\pdfoutput=1

\documentclass[onecollarge,natbib]{svjour2}
\bibpunct{[}{]}{;}{n}{}{,} 

\usepackage{graphicx}
\usepackage{hyperref}

\journalname{Few-Body Systems}

\begin{document}

\title{Generalized Valon Model for Double Parton Distributions \thanks{Talk presented by WB at the Light Cone 2015 Conference, Frascati, Italy, 21-25 September 2015.
Supported by Polish National Science Center (grant  DEC-2015/17/B/ST2/01838), by Spanish DGI (grant FIS2014-59386-P), and by Junta de Andaluc\'ia (grant FQM225).}
}

\titlerunning{dPDF in valon model}     

\author{Wojciech Broniowski \and Enrique Ruiz Arriola \and Krzysztof Golec-Biernat
}

\authorrunning{W. Broniowski et al.} 

\institute{W. Broniowski \at
             Institute of Nuclear Physics, Polish Academy of Sciences, 31-342 Cracow, Poland, and \\
             Institute of Physics, Jan Kochanowski University, 25-406 Kielce, Poland\\  
              \email{Wojciech Broniowski@ifj.edu.pl}  
                          \and
            E. Ruiz Arriola \at
              Departamento de F\'isica At\'omica, Molecular y Nuclear, and \\ Instituto Carlos I de F\'isica Te\'orica y Computacional, Universidad de Granada, E-18071 Granada, Spain \\
              \email{earriola@ugr.es}
                         \and
           K. Golec-Biernat \at
              Institute of Nuclear Physics Polish Academy of Sciences, 31-342 Cracow, Poland and \\
              Faculty of Mathematics and Natural Sciences, University of Rzesz\'ow, 35-959 Rzesz\'ow, Poland\\
              \email{golec@ifj.edu.pl}      
}

\date{Received: date / Accepted: date}

\maketitle

\begin{abstract}
We show how the double parton distributions may be obtained consistently from the many-body light-cone wave functions. We illustrate 
the method on the example of the pion with two Fock components. 
The procedure, by construction, satisfies the Gaunt-Stirling sum rules.
The resulting single parton distributions of valence quarks and gluons are consistent with a phenomenological parametrization at a low scale.
\end{abstract}

\vspace{5mm}

Multiparton distributions were considered already in pre-QCD times~\cite{Kuti:1971ph} and in the early days of QCD~\cite{Konishi:1979cb,Kirschner:1979im,Shelest:1982dg}. 
Recently, the interest in these objects has been renewed (see, e.g.,~\cite{Bartalini:2011jp,Snigirev:2011zz,Diehl:2011yj,Blok:2011bu,Blok:2013bpa,Kasemets:2014yna,Astalos:2015ivw}
and references therein), 
with expectations that the double parton 
distributions (dPDF's) of the nucleon may be accessible from the double parton scattering contribution 
in certain exclusive production channels at the LHC~\cite{d'Enterria:2012qx,Luszczak:2011zp}. Whereas dPDF's are well defined objects, their 
explicit construction, or parametrization that can be used phenomenologically, is difficult to achieve in a way where all the formal constraints are 
satisfied. In particular, the frequently used product ansatz or its modifications is inconsistent with the basic 
requirements of the Gaunt-Stirling (GS) sum rules~\cite{Gaunt:2009re,Gaunt:phd}.

On the model side, there are only a few explicit calculations in the literature: the studies in the MIT bag model~\cite{Chang:2012nw} or in the constituent quark
model~\cite{Rinaldi:2013vpa} implement relativity approximately, hence the support of the distributions extends outside of the physical 
region. These problems were mended in Refs.~\cite{Rinaldi:2014ddl,Rinaldi:2015cya}. On the 
other hand, a simple {\em valon} model~\cite{Hwa:1980mv,Hwa:2002mv} applied to the nucleon in~\cite{Broniowski:2013xba} leads to dPDF's satisfying 
all formal constraints and to reasonable single parton distributions (sPDF's) for the valence sector. In this talk we pursue this idea, extending the valon model to 
include higher Fock-state components, which allows us to study the gluon and sea content.

For practical purposes, it would be highly desirable to have a simple working parametrization for dPDF's in analogy to 
parameterizations of sPDF's. 
The ultimate goal of our approach is to systematically construct dPDF's which on one side satisfy all formal constraints, and on the 
other side reproduce the known sPDF's. 

In phenomenological applications one assumes for simplicity the transverse-longitudinal decoupling in the dPDF's, $\Gamma_{ij}(x_1,x_2,k_T) = D_{ij}(x_1,x_2) f(k_T)$, 
moreover, one frequently applies the ansatz
$D_{ij}(x_1,x_2) = D_i(x_1)D_j(x_2) \theta(1-x_1-x_2) (1-x_1-x_2)^n/(1-x_1)^{n_1}/(1-x_2)^{n_2}$, where $D_i(x)$ are the sPDF's. 
However, this form violates the GS sum rules~\cite{Gaunt:2009re}:
\begin{eqnarray}
&&\int_0^{1-y} \!\!\! dx \, D_{i_{\rm val} j}(x,y)=\int_0^{1-y} \!\!\! dx \, [D_{i j}(x,y)-D_{\bar{i}j}(x,y)] = 
(N_{i_{\rm val}}-\delta_{ij}+\delta_{\bar{i}j})D_j(y), \label{eq:gs} \\
&&\sum_{i} \int_0^{1-y} \!\!\! dx \,x D_{ij}(x,y)=(1-y) D_{j}(y), \;\;\; (\rm{and~similarly~for~the~second~parton}). \nonumber 
\end{eqnarray}
This violation is formally a serious problem, as sum rules~(\ref{eq:gs}) follow from very basic field-theoretic features, namely the 
Fock-space decomposition of the hadron wave function and conservation laws.
An attempt of a construction of the ansatz made in Ref.~\cite{Golec-Biernat:2014bva} met  problems with the parton exchange symmetry and positivity.
For the gluon sector of the proton, a formally successful ansatz has been obtained with the help of the Mellin moments~\cite{Golec-Biernat:2015aza}.

\begin{figure*}
\centering
\includegraphics[width=0.45\textwidth]{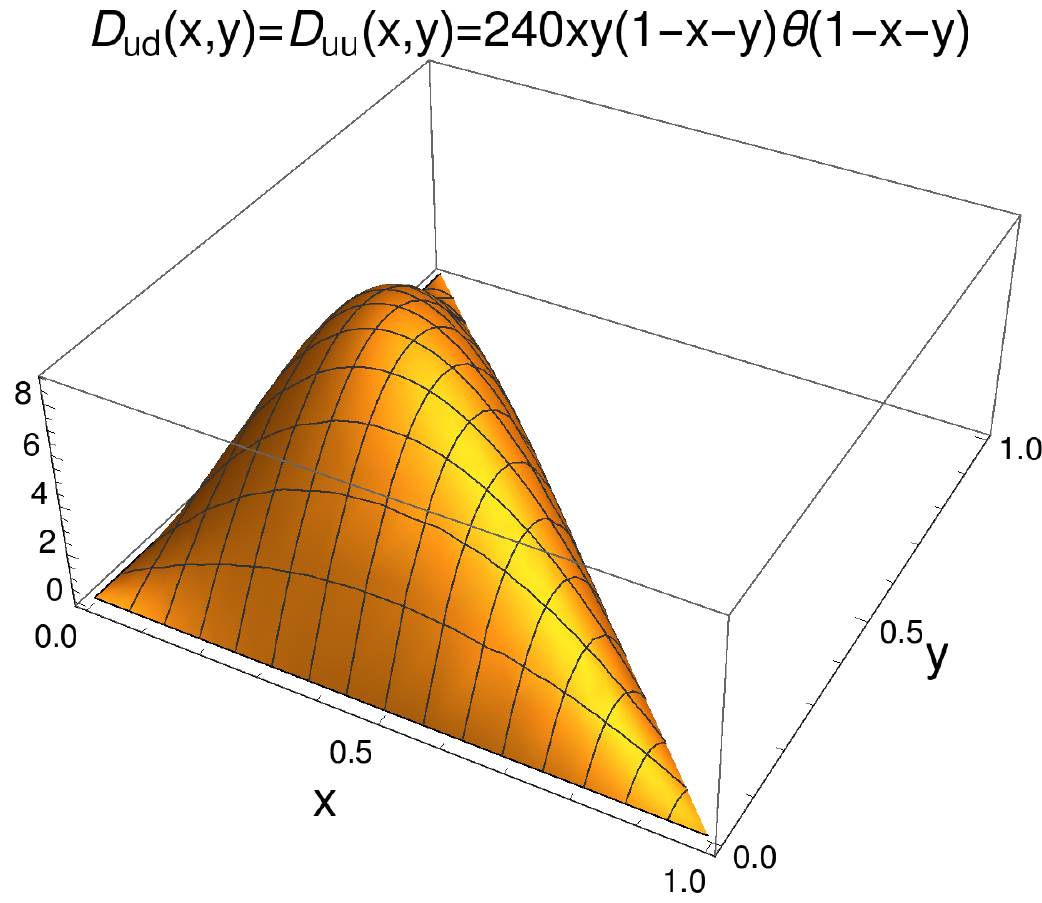} \hfill \includegraphics[width=0.39\textwidth]{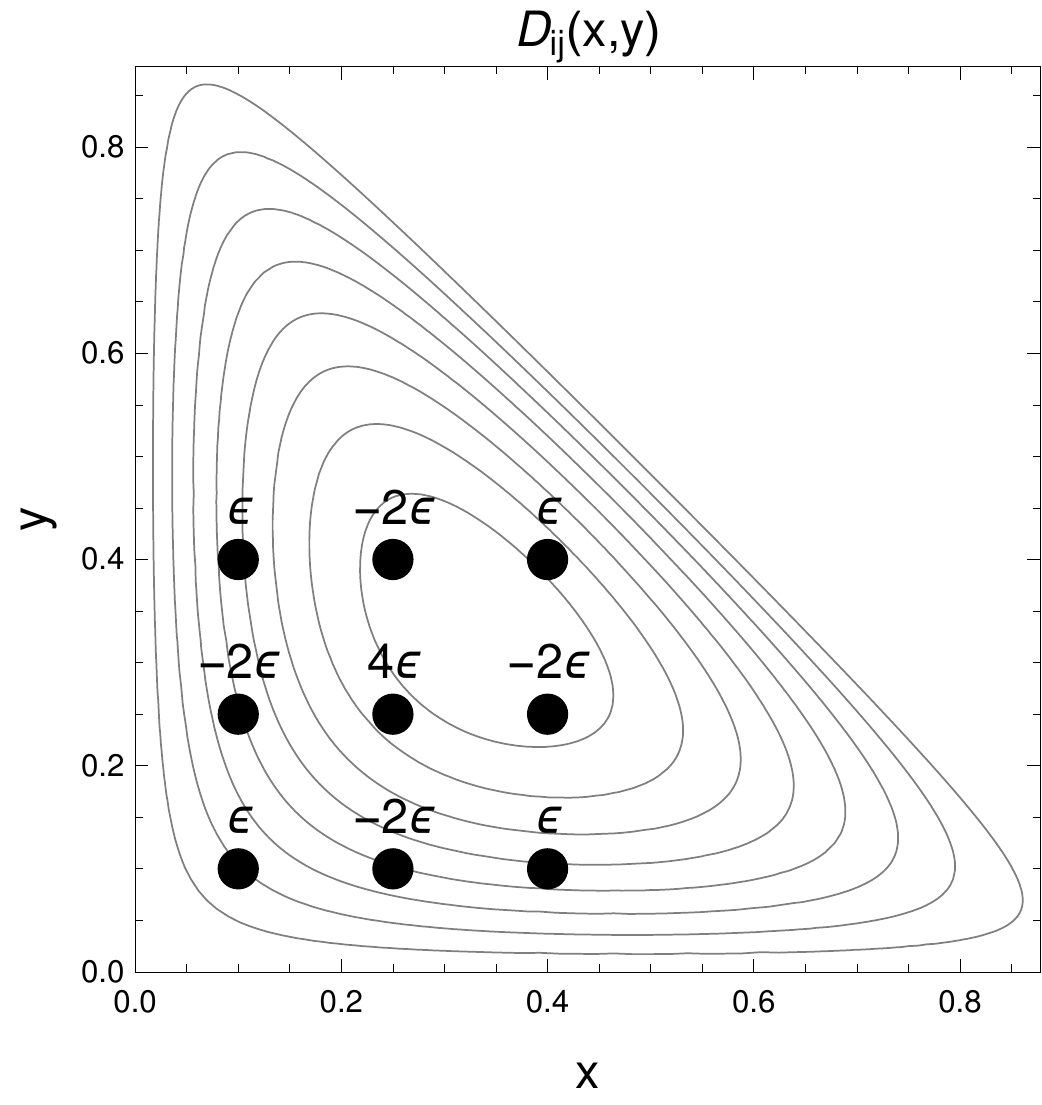}
\caption{Left: example of the valence dPDF of the proton which satisfies the GS sum rules. Right: schematic illustration of the non-uniqueness of the sPDF constraints.}
\label{fig:sr}     
\end{figure*}

The first issue we wish to point out in this talk is the non-uniqueness of the sPDF constraints. In other words, the sPDF's do not fix the dPDF's unambiguously.
Suppose we have found a form of dPDF's which satisfies the sum rules~(\ref{eq:gs}). 
A sample function illustrating the situation follows from the valon model for the nucleon~\cite{Broniowski:2013xba} with a single valence 
Fock component  $|p\rangle=|uud\rangle$. It is presented in the left panel of Fig.~\ref{fig:sr}. Projections lead to $D_u(x)=2D_d(x)=40x(1-x)^3$, where these 
valence sPDF's satisfy the GS sum rules, as can be explicitly seen:
\begin{eqnarray}
&& D_u(y)=\int dx \, D_{du}(x,y)=\int dx \, D_{uu}(x,y), \;\; 2D_d(y)=\int dx \, D_{ud}(x,y),   \\
&& (1-y) D_u(y)=\int dx \, x [D_{du}(x,y)+D_{uu}(x,y)], \;\;(1-y) D_d(y)=\int dx \, x D_{ud}(x,y). \nonumber
\end{eqnarray}
To show that the solution for dPDF's leading to specified sPDF's is not unique, we may perturb the dPDF's by adding or subtracting strength in specific points, as shown in the right panel 
of Fig.~\ref{fig:sr}. Then, by our specific choice of adding delta functions of strength $\epsilon$ and subtracting $-2\epsilon$ (or their multiplicities) at coordinates which are in proportion $2:5:8$ does not contribute to 
the right-hand sides of the sum rule. 
One may smear this construction by using a distribution of points. From a general point of view, the non-uniqueness is obvious: one-particle distributions do not fix the two-particle distribution, which may 
contain correlations~\cite{Snigirev:2014eua}. This shows that the ``bottom-up'' attempts to guess dPDF's with just the constraints from sPDF's are arbitrary.

As a remedy, we propose the ``top-down'' method which starts from the multiparticle light-cone wave function, extending the analysis of~\cite{Broniowski:2013xba}.
The approach guarantees the fulfillment of the formal requirement, in particular the GS sum rules~(\ref{eq:gs})~\cite{Gaunt:phd}.
The Fock expansion of a hadron state in partonic constituents is assumed to have the form
\begin{eqnarray}
|h\rangle= \sum_N\sum_{f_1\ldots f_N}\int dx_1\dots dx_N\,\delta(1-\sum_{k=1}^N x_k) \Psi_N(x_1\dots x_n;f_1\dots f_N)\,|x_1\ldots x_N;f_1\dots f_N\rangle, \label{eq:decom}
\end{eqnarray}
with $f_i$ denoting the parton type. One should then model the $N$-parton Fock components $\Psi_N$'s, obeying the constraints from the known sPDF's. 
Having this,  one may compute the double distributions, $D_{f_1f_2}(x_1,x_2)$, and also the higher-particle distributions if needed. 

We emphasize that the presented analysis based on Eq.~(\ref{eq:decom})
is effectively one-dimensional, with the transverse degrees of freedom integrated out. 
Non perturbative transverse lattice calculations suggest that such a 
dimensional reduction is
triggered at lower renormalization scales set by the lattice spacing, 
$Q \sim 1/a_\perp$.
Then, the transverse degrees of freedom
are effectively frozen  when the transverse momenta of the quarks are
smaller
than the lattice spacing~\cite{Burkardt:2001jg,Dalley:2002nj} (see also 
the discussion
in Ref.~\cite{Broniowski:2013xba}).

We make the simplifying assumption that the only correlations in the wave function come from the longitudinal momentum conservation: $1=x_1+x_2+\dots+x_n$ 
(the {\em generalized valon model}). Then
\begin{eqnarray}
&&{|\psi_N(x_1\dots x_n;f_1\dots f_N|^2 = A^2 \phi_{f_1}(x_1)\dots\phi_{f_N}(x_N)}, 
\end{eqnarray}
where $\phi_{f_1}(x)=|\psi_{f_i}(x)|^2$ and $\psi_{f_i}(x)$ is the one-body wave function of parton $f_i$, in principle computable in a dynamical model.
Let the asymptotics of the single-parton functions be
\begin{eqnarray}
&& \phi_{f_i}(x) \sim x^{\alpha_{f_i}-1} \;\;\; {\rm at}~ x \to 0, \;\;\;\;\;\; \phi_{f_i}(x) \sim (1-x)^{\beta_{f_i}} \;\;\; {\rm at}~ x \to 1,
\end{eqnarray}
where for integrability $\alpha_{f_i}>0$ and  $\beta_{f_i}>-1$. 
For $N=2$ with parton types $f$ and $f'$ the resulting asymptotics for sPDF's  
at $x\to 0$ and $x\to 1$  is $D_{f}(x) \sim x^{\alpha_{f}+\beta_{f'}-1}$ and  $D_{f}(x) \sim (1-x)^{\beta_{f}+\alpha_{f'}-1}$, 
whereas dPDF's assume the singular form  $D_{ff'}(x,y) \sim \phi_f(x) \phi_{f'}(y) \delta(1-x-y)$.
The emergence of the singular part follows from the presence of just two partons. It would be washed out by the QCD evolution to higher scales~\cite{Broniowski:2013xba}, 
which redistributes the strength to higher Fock components. For $N \ge 3$ 
\begin{eqnarray}
&& D_{f}(x) \sim x^{\alpha_{f}-1} \;\;\; {\rm at}~ x \to 0, \;\;\;\;\; {D_{f}(x) \sim (1-x)^{\beta_{f}+\alpha_{f_{(1}}+ \dots + \alpha_{f_{N-1)'}} -1}  \;\;\; {\rm at}~ x \to 1}, \label{eq:asym}
\end{eqnarray}
where $(...)'$ indicates that index $f$ is omitted in the sequence. 
We note that the $x\to 1$ behavior is sensitive to the low-$x$ behavior of the other components, as in this limit the kinematics 
``pushes'' them towards 0. The above formulas are useful in modeling, as they allow for matching of the asymptotic behavior of the single-particle functions $\phi_{f_i}$ to the 
phenomenologically known asymptotics of sPDF's.

Before we apply the extended valon model to a specific case, let us make some remarks on the choice of the scale and the QCD evolution.
With an increasing scale $Q_0$, more and more partons at low $x$ are generated, hence more and more Fock components are needed in the decomposition (\ref{eq:decom}).
For practical reasons of not dealing with too many components, it is then favorable to use the parameterizations for sPDF's at a lowest possible $Q_0$, such as, e.g., 
the GRV$_\pi$~\cite{Gluck:1991ey} parametrization for the pion which uses $Q_0$ as low as 500~MeV.
The tempting evolution to lower scales cannot be carried out too far down, as negative distributions are generated~\cite{LlewellynSmith:1978me,RuizArriola:1998er}, as well as non-perturbative domain 
is entered. We also remark that the constituent quark models which do not have gluons or quarks at the {\em quark model scale}~\cite{Davidson:1994uv,Broniowski:2007si} 
do not generate sufficiently many gluons and sea quarks at higher experimental scales.

Explicitly, the GRV$_\pi$~\cite{Gluck:1991ey} parametrization for the valence, gluon and sea sPDF's reads 
\begin{eqnarray}
 && x V(x) = 0.52 \left(0.38 \sqrt{x}+1\right) (1-x)^{0.37} x^{0.50}, \label{eq:grv}\\
 && x g(x) = \left(0.34 \sqrt{x}+0.68\right) (1-x)^{0.39} x^{0.48}, \;\; x q_{\rm sea}(x)=0, \;\;\; (Q_0=500 {\rm MeV}). \nonumber
\end{eqnarray}
In the following we will use the momentum fractions as constraints:
\begin{eqnarray}
&& 2\int dx\, x V(x) = 0.58, \;\;\; \int dx\, x g(x) =0.42. 
\end{eqnarray}
The average number of gluons inferred from Eq.~(\ref{eq:grv}) is $\int dx\, g(x)= 1.46$, hence a component with at least two gluons is necessary in the wave function.

We use the simple ansatz with just two Fock components:
\begin{eqnarray}
|\pi^+\rangle = A |\bar{u}d\rangle +B  |\bar{u}d gg \rangle. \label{eq:two}
\end{eqnarray}
Of course, this is a simplification, as we could have components with a single gluon and more than two gluons, as well as quark sea contributions.
The corresponding sPDF's are:
\begin{eqnarray}
&&  D_{\bar u}(x)=A^2 |\Psi_{\bar{u}d}(x,1-x)|^2+ B^2\! \int \!dx_3dx_4  |\Psi_{\bar{u}dgg}(x,1-x-x_3-x_4,x_3,x_4)|^2 = D_{d}(x),   \\
&&  D_{g}(x)= B^2\! \int \!dx_1dx_2  \left ( |\Psi_{\bar{u}dgg}(x_1,x_2,x,1-x-x_1-x_2)|^2 + |\Psi_{\bar{u}dgg}(x_1,x_2,1-x-x_1-x_2,x)|^2 \right ). \nonumber 
\end{eqnarray}
The conditions
\begin{eqnarray}
&& A^2+B^2=1, \label{eq:2cond}\\
&& \int dx\,x [ D_{\bar u}(x)+ D_{d}(x)]=0.58 \;\;\; {\rm or} \;\;\; \int dx\,x  D_{g}(x)=0.42 \nonumber
\end{eqnarray}
provide constraints for parameters. We use the generalized valon ansatz
\begin{eqnarray}
&& |\Psi_{\bar{u}d}(x_1,x_2)|^2  \sim  f(x_1; a,b) f(x_2;a,b),  \label{eq:psi}\\
&& |\Psi_{\bar{u}dgg}(x_1,x_2,x_3,x_4)|^2 \sim f(x_1;\alpha_q,\beta_q) f(x_2;\alpha_q,\beta_q) f(x_3;\alpha_g,\beta_g) f(x_4;\alpha_g,\beta_g) \nonumber
\end{eqnarray}
with $f(x;\alpha,\beta)=x^{\alpha -1} (1-x)^{\beta}$.
The choice $a+b=0.5$, $\alpha_q=0.5$, $\beta_q=-0.09$, $\alpha_g=0.48$, $\beta_g=-0.09$ is consistent with the asymptotic limit (\ref{eq:asym}). Application of 
(\ref{eq:2cond}) yields $A^2=0.15$ and $B^2=0.85$, which means a strong dominance of the component with gluons over the $q\bar{q}$ component in the considered model. 

\begin{figure*}
\centering
\includegraphics[width=0.47\textwidth]{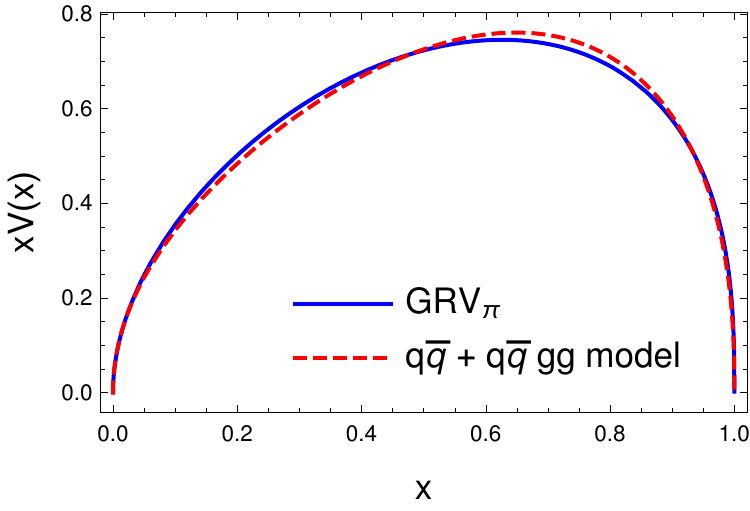} \hfill \includegraphics[width=0.47\textwidth]{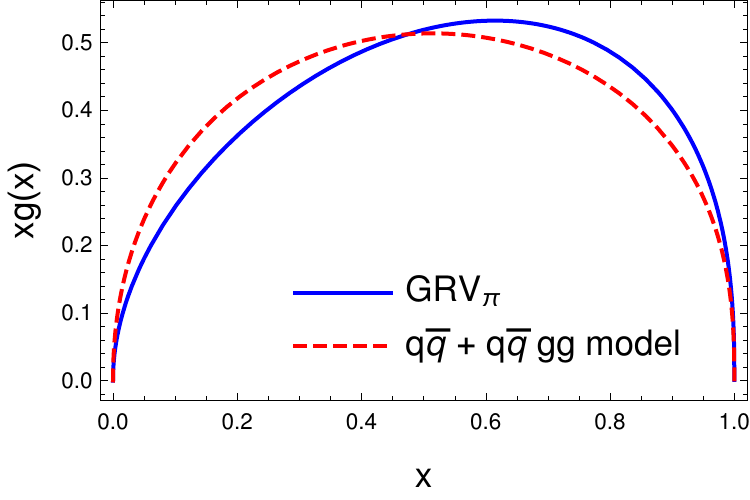}
\caption{The valence quark (left) and gluon (right) sPDF's of the pion from the two-component valon model (\ref{eq:two},\ref{eq:psi}) compared to the the GRV$_\pi$ parametrization~\cite{Gluck:1991ey}.}
\label{fig:single}   
\end{figure*}

\begin{figure*}
\centering
\includegraphics[width=0.44\textwidth]{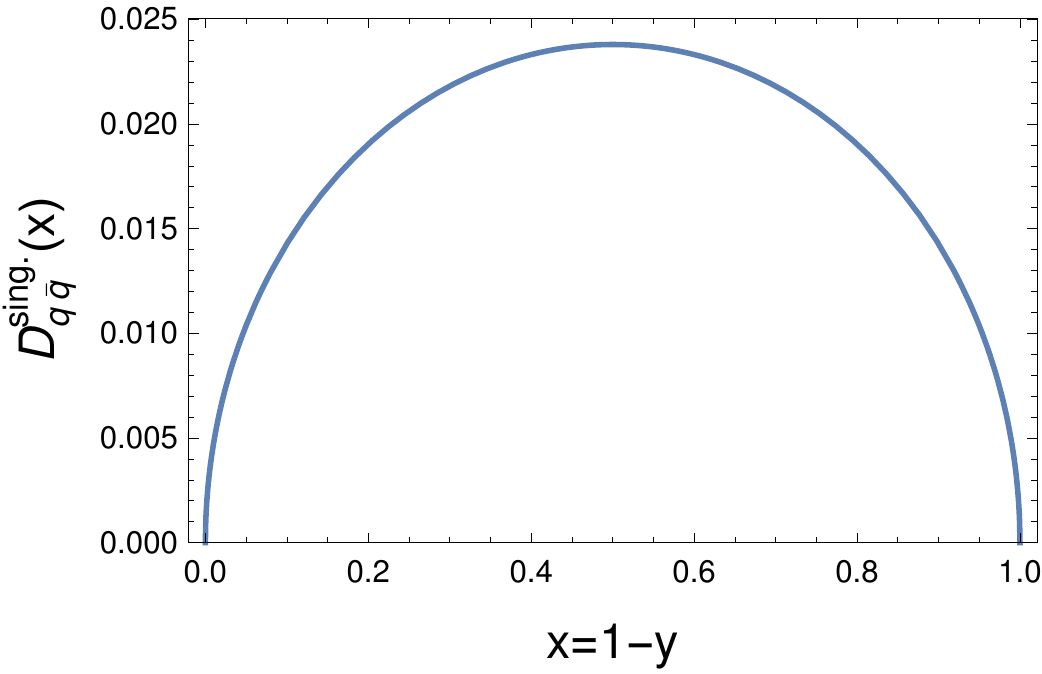}\hfill \includegraphics[width=0.44\textwidth]{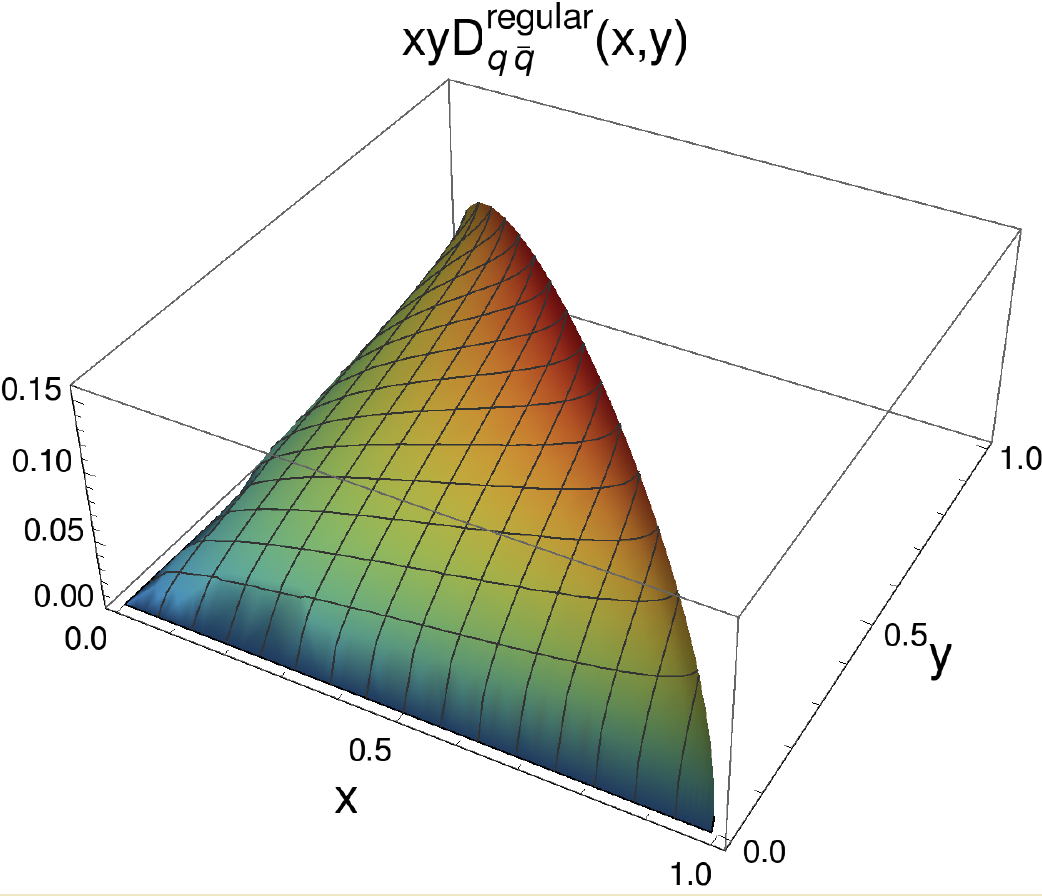} \\ ~ \\
\includegraphics[width=0.44\textwidth]{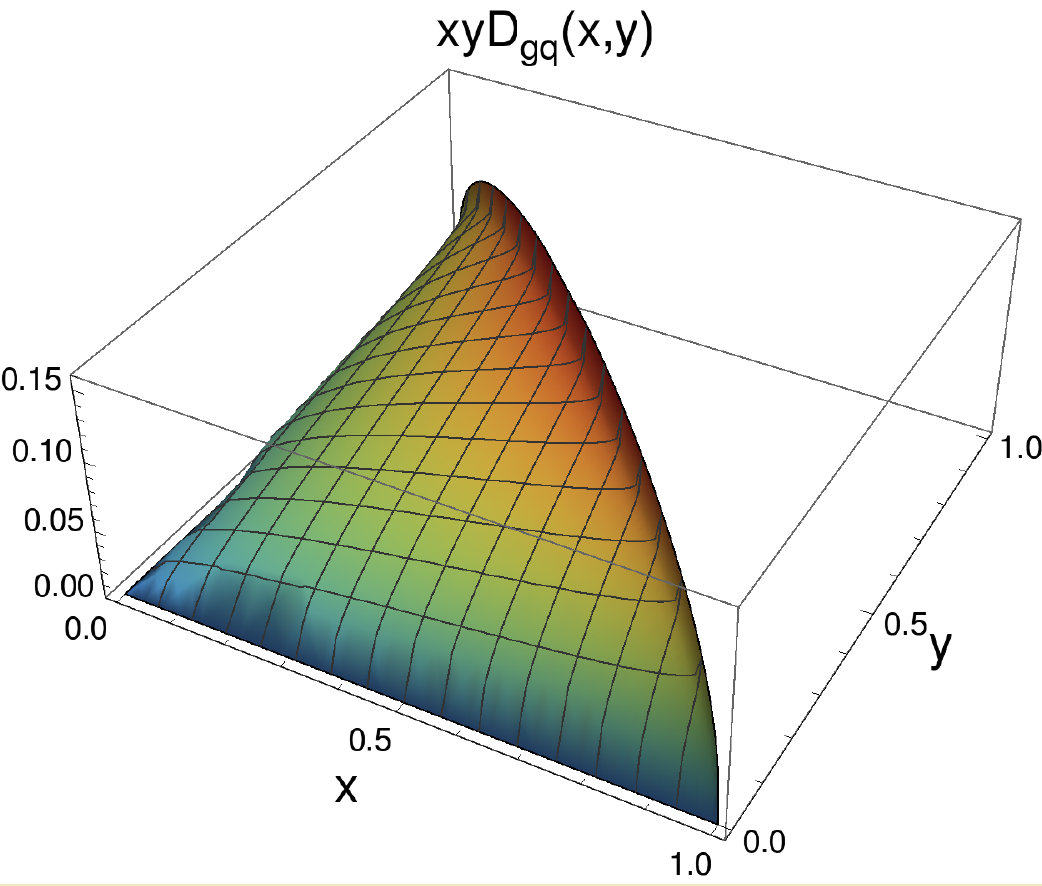}\hfill \includegraphics[width=0.44\textwidth]{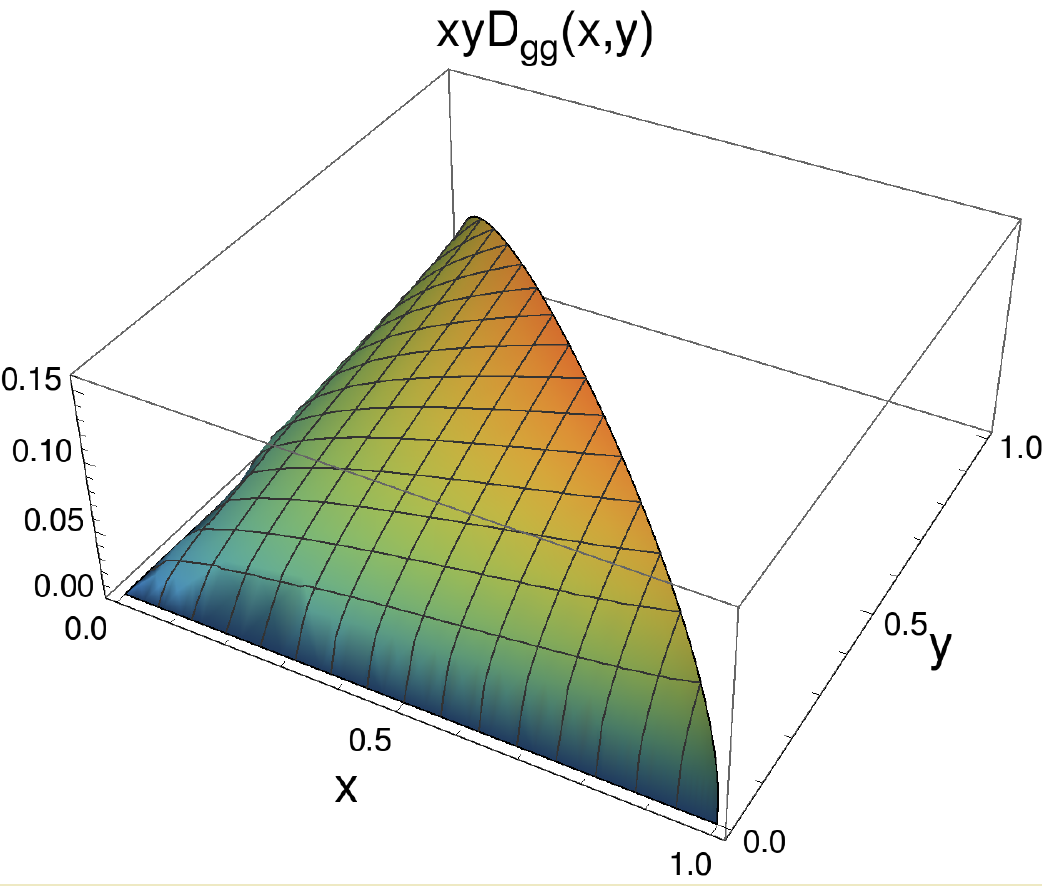}
\caption{dPDF's of the pion from the the two-component valon model  (\ref{eq:two},\ref{eq:psi}).}
\label{fig:double}   
\end{figure*}

The resulting sPDF's are shown in Fig.~\ref{fig:single}. We note a quite good (taking into account the simplicity of the model) agreement with the GRV$_\pi$ parametrization, especially for the 
valence quarks. We can now tackle with our main task, namely, the determination of dPDF's. 
As already mentioned, the $N=2$ component generates a singular part $D_{\bar{u} d} \sim f(x;a,b) f(y;a,b)\delta(1-x-y)=D_{\bar{u} d}^{\rm sing}(x,y) \delta(1-x-y)$. 
The $N=4$ component leads, upon integration over $x_3$ and $x_4$, to 
broadly distributed functions. The result is shown in Fig.~\ref{fig:double}, proving that the proposed construction is practical.

An analogous construction for the case of the nucleon would be more challenging, as it requires at least three components 
in the Fock-space decomposition:
\begin{eqnarray}
|p\rangle = A_{uud} |uud\rangle +\dots + A_{uud\bar{q}q}   |uudq\bar{q}\rangle +\dots + A_{uudgg}   |uudgg\rangle
\end{eqnarray}
This study, which would generalize the results for the valon model $|p\rangle = |uud\rangle$ considered in~\cite{Broniowski:2013xba}, is left for the future. 

In conclusion, here are our main points:

\begin{itemize}
  
   \item The top-down strategy of constructing multi-parton distributions, which guarantees the  formal features, is practical when the number of the Fock components is not too large.
   This is the case of dynamics at the  lowest possible scale.
   
   \item The approach requires modeling of the light-cone wave functions, hence has physical input.
   
   \item Phenomenological sPDF's are used as constraints, but sPDF's alone cannot uniquely fix dPDF's.
   
    \item Many Fock components are needed to accurately reproduce  the popular parameterizations of sPDF's, even at relatively low scales. The QCD evolution takes care of the generation 
    of the higher Fock components at higher momentum scales.

    \item The valon model offers a simple ansatz at the initial low-energy scale that grasps the essential features with just the longitudinal momentum conservation, and the 
      transverse degrees of freedom separated out from the dynamics.

    \item We note that the QCD evolution washes out the correlations in dPDF's at low $x_1$ and $x_2$, justifying the approximate validity of the product ansatz in that limit. However, outside of that 
       region the correlations are substantial.
    
\end{itemize}



\end{document}